\documentclass[a4paper,12pt]{article}

\usepackage{amsmath,amssymb,array,amsfonts}
\usepackage[dvipdfmx]{graphicx}
\usepackage{comment,xcolor}

\allowdisplaybreaks

\makeatletter
    
    \@addtoreset{equation}{section}
\makeatother

\usepackage[dvipdfmx]{hyperref} % hyper link
\hypersetup{%draft = true,%
			linkcolor= purple,%
			colorlinks= true,
			citecolor = cyan
			}

\def\e {{\rm e}}	
\def\s {{\rm s}}
\def\d {{\rm d}}

\newcommand{\bs}{\boldsymbol}

\newcommand{\mmot}{\!\!\not \!\! }

\newcommand{\beq}{\begin{equation}}
\newcommand{\eeq}{\end{equation}}
\newcommand{\bea}{\begin{eqnarray}}
\newcommand{\eea}{\end{eqnarray}}
\newcommand{\dis}{\displaystyle}

\newcommand{\nt}{\notag}

\begin{document}
\setlength{\baselineskip}{18pt}
\begin{titlepage}

\begin{flushright}
OCU-PHYS 394
\end{flushright}
\vspace{1.0cm}
\begin{center}
{\Large\bf 
$\mu \to 3e$ and $\mu \to e$ Conversion %Processes \\
%\vspace*{3mm}
in Gauge-Higgs \\ 
\vspace*{3mm}
Unification 
} 　
\end{center}
\vspace{25mm}

\begin{center}
{\large
Yuki Adachi, 
%\footnote{e-mail : y-adachi@matsue-ct.ac.jp},
%
Nobuaki Kurahashi$^*$, 
%\footnote{e-mail : 075s112s@stu.kobe-u.ac.jp},
%
and Nobuhito Maru$^{**}$
%\footnote{e-mail : maru@phys-h.keio.ac.jp}
%
}
\end{center}
\vspace{1cm}
\centerline{{\it
Department of Sciences, Matsue College of Technology,
Matsue 690-8518, Japan.}}

\centerline{{\it
$^*$Department of Physics, Kobe University,
Kobe 657-8501, Japan.}}

\centerline{{\it
$^{**}$Department of Mathematics and Physics,
%}}
%\centerline{{\it
Osaka City University,
Osaka 558-8585, Japan.
}}

%%%%%%%%%%%%%%%%%%%%%%%%%%%%%%% Abstract %%%%%%%%%%%%%%%%%%%%%%%%%%%%%%%
%
\vspace{0.5cm}

\begin{abstract}
The two dominant processes of lepton flavor violation, 
 $\mu \to 3e $ and $\mu \to e$ conversion in atomic nuclei 
 caused by neutral KK gauge boson exchanges at tree level, 
 are studied in the context of five dimensional gauge-Higgs unification scenario. 
The key point of the flavor violation is a fact that the bulk masses and the brane ones cannot be simultaneously diagonalized. 
We estimate the branching ratio of each processes
 and obtain the lower bound of compactification scale around weak scale from the current experimental data. 
We discuss the reasons why the final result is not so severe 
 although the large mixing in the lepton sector seems to give large lepton flavor violation processes.  
\end{abstract}

\end{titlepage}

%\maketitle
%\thispagestyle{empty}

%\vspace{5mm}

%}

\newpage
%%%%%%%%%%%%%%%%%%%%%%%%%%%%%%% Introduction %%%%%%%%%%%%%%%%%%%%%%%%%%%%%%%
\section{Introduction}
It was a great triumph that a Higgs boson was discovered 
 by ATLAS \cite{ATLAS} and CMS \cite{CMS} collaborations at the CERN LHC, 
 but it is still unclear that the observed Higgs boson is a predicted one in the Standard Model (SM) or the physics beyond the SM.   
Of the various physics beyond the SM, 
 the gauge-Higgs unification (GHU) \cite{GH} is one of the fascinating scenarios as a solution to the hierarchy problem \cite{HIL}. 
In GHU, Higgs boson is identified with extra spatial components of the gauge boson in higher dimensional gauge theories. 
This identification immediately forbids a local mass term for Higgs boson by the higher dimensional gauge symmetry, 
 which makes us expect quantum corrections to Higgs mass to be finite in compactified theories. 
However, we have to check the expectations explicitly since the theories under consideration are non-renormalizable. 
In fact, the finiteness of Higgs mass has been checked so far in various types of compactifications \cite{HIL, examples} 
 and up to 2-loops of perturbations \cite{2loop}. 
The finite observables other than Higgs mass and potential have been also known so far, 
 such as the gluon fusion production and diphoton decay of Higgs boson \cite{Maru},  the g-2 and EDM \cite{g-2, EDM}, 
 and the deviation of Yukawa coupling from the gauge coupling \cite{LM}.

The observables in the gauge and Higgs sector of the theory 
 are well controlled by the gauge symmetry and predictable, 
 but a matter sector is troublesome in GHU 
 because of a feature such that Yukawa coupling is originated from the gauge coupling.  
Therefore, issues of Yukawa hierarchy, flavor mixings and CP violation are very nontrivial in GHU. 
As for Yukawa hierarchy, the fermion mass is provided by W-boson mass to start with. 
The light fermion Yukawa couplings are realized by the overlap integral of wave functions localized 
 at different point in extra space.   
Changing the location of the wave functions is realized by $Z_2$ odd bulk mass parameters 
 in a five dimensional case considered in this paper. 
A realistic Yukawa couplings except for top quark is obtained 
 by ${\cal O}(1)$ tuning of the $Z_2$ odd bulk mass parameters. 
Top yukawa coupling is realized by embedding to the large dimensional representation. 
The mechanism of generating the flavor mixing in GHU was proposed in \cite{BN, KK}, 
 and was applied to the FCNC processes in the quark sector \cite{KK, DD, BB}.  
To obtain flavor mixings, both the bulk masses and the brane ones are necessary, 
 where the bulk masses are for Yukawa hierarchy as mentioned before 
 and the brane masses are for removing massless exotic fermions as explained later. 
In general, these bulk masses and the brane ones cannot be simultaneously diagonalized, 
 which results in the flavor mixing. 
As for CP violation in the context of GHU,  
 some mechanisms have been proposed, 
 where CP is spontaneously broken by the vacuum expectation value (VEV) of 
 CP odd extra component of higher dimensional gauge boson in odd extra dimensions \cite{EDM}, 
 by compactification with a complex structure in even dimensions \cite{LMN} 
 and by the phases in the flavor mixing matrices for quark doublets \cite{AKMT}.  

In this paper, we study the lepton flavor violation in the context of GHU. 
Similar to the quark sector, the lepton flavor violation appears 
 in the presence of both the bulk masses and the brane ones and 
 by nonzero Kaluza-Klein (KK) Z boson and photon exchange at tree level. 
We have a naive guess that the lepton flavor violation is larger than the flavor mixing in the quark sector 
 because of the large mixing in the lepton sector contrary to the small mixing in the quark sector. 
 We therefore expect that the lepton flavor violation processes put more severe constraints for parameters of the theory.   
As typical examples, we consider two processes in the context of GHU, namely $\mu \to 3e$ and $\mu \to e$ conversion in nuclei. 

This paper is organized as follows. 
We introduce our model and explain how the lepton flavor mixings are obtained in  the next section. 
In section 3, we estimate the processes of $\mu \to 3e$ and $\mu \to e$ conversion in nuclei 
 and derive the lower bound for the compactification scale from the experimental data.
Summary is given in section 4. 
In appendix \ref{modelparameter}, 
we provide the model parameters used in the calculation.

%%%%%%%%%%%%%%%%%%%%%%%%%%%%%%% Model %%%%%%%%%%%%%%%%%%%%%%%%%%%%%%%
\section{Model}
\label{Themodel}
We consider a five dimensional $SU(3) \times U'(1) \times SU(3)_\text{color}$
 gauge-Higgs unification model compactified on an orbifold $S^1/Z_2$ with a compactification radius $R$. 
The electroweak symmetry $SU(2)_L \times U(1)_Y$ is embedded in $SU(3) \times U(1)'$ gauge group 
 and $SU(3)_\text{color}$ is a QCD color group. 
Since the quark sector of the model was discussed in detail 
 by the present authors in \cite{KK, DD, BB}, 
 we briefly summarize our model.

The leptons and quarks except for the top quark are embedded in $\bf 3$ and $\bf \bar 6$ representations of $SU(3)$,
 and the top quark only needs a higher dimensional representation such as the $\bf \overline{15}$ of $SU(3)$ 
 for the following reason \cite{CCP}.
In the gauge-Higgs unification scenario, 
 the yukawa couplings are originated from the higher-dimensional gauge coupling 
 so that the various light fermion masses are realized by the overlap integral of wave functions for quarks and leptons, 
 in which the wave function is localized at different point in the fifth dimension. 
 %by the localized profiles in our model. 
On the other hand, the top quark mass around twice of the W boson mass is realized 
 by the normalization factor of higher dimensional representation. 
The matter content in our model is summarized as
\begin{align}
 q({\bf 3},0,{\bf 3})^i
=&
  Q_3^i+d^i~~~~~(i=1,2,3)
  ,
  \\
 q({\bf \bar 6},0,{\bf 3})^i
\supset&
 Q_6^i+u^i ~~~~~(i=1,2)
 ,
 \\
 q({\bf \overline{15}},-2/3,{\bf 3})
\supset &
 Q_{15}+t, 
 \\
  l({\bf 3},-2/3,{\bf 1})^i
=&
  L_3^i+e^i~~~~~(i=1,2,3)
  ,
  \\
 l({\bf \bar 6},-2/3,{\bf 1})^i
\supset&
 L_6^i+\nu^i ~~~~~(i=1,2,3),
\end{align}
where the numbers in the parenthesis stand for  the representations of $SU(3), U'(1)$ and $SU(3)_{\text{color}}$. 
The $Q$ and $L$ represent the quark and lepton $SU(2)_L$ doublets, respectively and
the $u, d, t, \nu$ and $e$ represent the $SU(2)_L$ singlets of the quarks and leptons. 
$\supset$ means that the representations except for the fundamental representation have fields irrelevant for the SM fermions. 
The $U'(1)$ charges of each representations are adjust for each SM fields to have appropriate hypercharges. 

Then the bulk Lagrangian consists of three parts; 
\bea
\mathcal L_\text{B} = \mathcal L_\text{q} + \mathcal L_\text{l} + \mathcal L_\text{G}.
\eea
Each Lagrangians are given as follows;
\begin{subequations}
\begin{align}
 \mathcal L_\text{q}
=&
 \bar q_3 ^i ( i\mmot D_3 - M^i_q \epsilon(y)) q_3^{i} 
  +\bar q_{\bar 6}^j ( i\mmot D_{\bar 6} -M^j_q \epsilon(y) ) q _{\bar 6}^j
  +\bar q_{\bar 6}^{j=3}  i\mmot D_{\bar 6} q _{\bar 6}^{j=3}
  +\bar q_{\overline {15}}  i\mmot D_{\overline {15}}'  q _{\overline {15}},
 \\
\mathcal L_\text{l}
=&
 \bar l_3 ^i ( i\mmot D_3' - M^i_l \epsilon(y)) l_3^{i} 
  +\bar l_{\bar 6}^i ( i\mmot D_{\bar 6}' -M^j_l \epsilon(y) ) l _{\bar 6}^i,
 \\
 \mathcal L_\text{G}
=&
 -\frac{1}{2}{\rm Tr}G_{MN}G^{MN}
 -\frac{1}{4}B_{MN}B^{MN} -\frac{1}{2}{\rm Tr} F_{MN}F^{MN}
  ,
\end{align}
\end{subequations}
where the index $i,j$ denotes a generation: $i=1,2,3$ and $j=1,2$.
The covariant devatives and the field strength of gauge bosons are defined by
\begin{subequations}
\begin{align}
\mmot D =& \Gamma^M (\partial_M -ig A_M -ig_\text{s} G_M),\\
\mmot D' =& \Gamma^M (\partial_M -ig A_M -ig' B_M -ig_\text{s} G_M),\\
	G_{MN}
 &=	\partial_MG_N-\partial_NG_M-ig_\s\big[G_M,G_N\big] 
 \ ,\\
	B_{MN}
 &=	\partial_MB_N-\partial_NB_M
 \ ,\\
	F_{MN}
 &=	\partial_MA_N-\partial_NA_M-ig\big[A_M,A_N\big]
 .
\end{align}
\end{subequations}
The $G_M,B_M$ and $A_M$ stand for the $SU(3)_\text{color},U'(1)$ and $SU(3)$ gauge fields respectively  and 
 the gauge fields $G_M$ and $A_M$ are written in a matrix form,
 e.g. $A_M = A_M^a \frac{\lambda^a}2$ in terms of Gell-Mann matrices $\lambda^{a}$. 
$M,N=0,1,2,3,5$ and the five dimensional gamma matrices
 are given by $\Gamma^M=(\gamma^{\mu},i\gamma^{5})$ ($\mu=0,1,2,3$). 
$g_\text{s}, g$ and $g'$ are 5D gauge coupling constants of $SU(3)_\text{color},SU(3)$ and $U'(1)$, respectively. 
$M^i$ are generation dependent bulk mass parameters of the fermions
accompanied by the sign function $\epsilon (y)$ with respect to the fifth dimensional coordinate $y$. 
Note that the  gauge fixing terms and ghost fields are needed to describe our model, however, we omitted them to avoid complexity.

The theory has a periodicity along with the $y$-direction 
 and $Z_2$ parities are assigned on the fields.
We adopt the $Z_2$ parities for the gauge fields
\begin{subequations}
\begin{align}
&G_\mu(-y) = G_\mu(y), \hspace*{1.5cm} G_y(-y) = - G_y(y), \\ 
&B_\mu(-y) = B_\mu(y), \hspace*{1.5cm} B_y(-y) = - B_y(y), \\
&A_\mu(-y) = P A_\mu(y) P^{-1}, \quad A_y(-y) = - P A_y(y) P^{-1}, 
\end{align}
where the orbifolding matrix is defined as $P={\rm diag}(-,-,+)$ and operated in the same way at the fixed points $y=0, \pi R$. 
For the matter fields, the parity assignments is as follows. 
\begin{align}
	q^i({\bs3})
 =&~\big\{Q_{3L}^i(+,+)+Q_{3R}^i(-,-)\big\}
	\oplus\big\{d_L^i(-,-)+d_R^i(+,+)\big\}
  \quad \big(\,i = 1,2,3\,\big)\ ,\\
	q^i(\bar{\bs6})
 \supset&~
	\big\{Q_{6L}^i(+,+)+Q_{6R}^i(-,-)\big\}
  ~\oplus\big\{u^i_L(-,-)+u^i_R(+,+)\big\}
  \quad \big(\,i = 1,2\,\big)\ ,\\
	q(\overline{\bf 15})
 \supset&~
	\big\{Q_{15L}(+,+)+Q_{15R}(-,-)\big\}
  \oplus\big\{t_L(-,-)+t_R(+,+)\big\} 
  \\
	l^i({\bs3})
 =&~\big\{L_{3L}^i(+,+)+L_{3R}^i(-,-)\big\}
	\oplus\big\{e_L^i(-,-)+e_R^i(+,+)\big\}
  \quad \big(\,i = 1,2,3\,\big)\ ,\\
	l^i(\bar{\bs6})
 \supset&~
	\big\{L_{6L}^i(+,+)+L_{6R}^i(-,-)\big\}
  ~\oplus\big\{\nu^i_L(-,-)+\nu^i_R(+,+)\big\}
  \quad \big(\,i = 1,2,3\,\big)\ ,
\end{align}
\end{subequations}
where the sign in the parenthesis represents the eigenvalues of the $Z_2$ parities at $y=0$ and $\pi R$, respectively.
Here we can see that the gauge symmetry $SU(3)$ is explicitly broken into $SU(2) \times U(1)$ and 
a chiral theory is realized in the zero mode sector by $Z_2$ orbifolding.

The mode functions of gauge fields are easily obtained with respect to the $Z_2$ parities.
The $Z_2$ even (odd) components are expanded in terms of 
 $C_n(y) = \frac1{\sqrt{\pi R}}\cos\frac nRy, S_n(y) = \frac1{\sqrt{\pi R}} \sin \frac nRy$.
Here we concentrate on the zero mode sector which plays an important role on the flavor mixing.
The five dimensional fermion $\chi^i$ ($\phi^i$) with $Z_2$ even (odd) parity includes the left- (right-) handed chiral fermions as
\begin{equation}
 \chi ^i \supset \chi_L^{i(0)}(x) f_L^i (y)\ , 
 \phi ^i \supset \phi_R^{i(0)}(x) f_R^i (y)\ ,
\end{equation}
where 
\begin{equation}
f_L^i (y) = \sqrt{\frac{M^i}{1-\e^{-2\pi RM^i}}} \e^{-M^i |y|}\ ,\ 
f_R^i (y) = \sqrt{\frac{M^i}{\e^{2\pi RM^i}-1}} \e^{M^i |y|}\ . 
\end{equation}
The above mode functions lead to the light fermion mass by strong suppressions for Yukawa couplings.

Since the $U(1)$ gauge boson includes a gauge anomaly in this model, 
 we discuss the mixing between $U(1)$ and $U'(1)$ gauge bosons.
The mixing between them has been already discussed in \cite{SSS},
 a hypercharge $U(1)_Y$ is identified with some linear combination of $U(1)$ and $U'(1)$ gauge groups.
The other orthogonal $U(1)$ is anomalous, and therefore the corresponding gauge boson obtain a large mass around the cutoff scale $\Lambda$.
Thus, the original $U(1)$ gauge bosons are separated into an anomalous one $Z'$ and the hypercharge gauge boson $A_Y$ as follows:
\begin{align}
 U(1)~\text{part}
%=&
% \frac{g}{2}\lambda_a A^a + g' Y' G\nt \\
\supset&
 \frac{g}{2}(\lambda _3A^3 +  \lambda_8 A^8)+g' Y' B \nt \\
=&
 \frac{g}{2}\lambda_3 A^3 +\left(\frac{g}{2}\cos\theta \lambda_8 -g'\sin\theta Y'\right)A_Y
 +\left(\frac{g}{2}\sin\theta \lambda_8 +g'\cos\theta Y'\right)Z',
\end{align}
where $A_Y,Z'$ are defined as follows:
\begin{equation}
\begin{cases}
A^8&=\cos\theta A_Y+\sin\theta Z',\\
B &=\cos\theta Z'-\sin\theta A_Y.
\end{cases}
\end{equation}
We assign the $U'(1)$ charge $-2/3$ on $\bf 3,\bf \bar 6$ 
 and the $U(1)_Y$ hypercharge is identified with a sum of $U(1)$ and $U'(1)$ charges. 
In this case, the $U(1)$, $U'(1)$, and $U(1)_Y$ charges are provided as
\begin{equation}
\lambda_8=\frac{1}{\sqrt{3}}{\rm diag}(1,1,-2),
Y'={\rm diag}\left( -\frac{2}{3},-\frac{2}{3},-\frac{2}{3} \right),
Y={\rm diag} \left( -\frac{1}{2},-\frac{1}{2},-1 \right), 
\end{equation}
from which we find
\begin{equation}
g_YY=\frac{g}{2}\cos\theta \lambda_8-g'\sin\theta Y',
\end{equation}
and
\begin{equation}
\cos\theta= \frac{g'}{\sqrt{3g^2+g'^2}}, \quad
\sin\theta = -\frac{\sqrt{3}g}{\sqrt{3g^2+g'^2}}, \quad 
g_Y = \sqrt{3}g \cos\theta.
\end{equation}
These results tell us photon and $Z$-bosons like 
\begin{align}
  \text{U(1) part}
=&
  \left(\frac{g}{2}\cos\theta_W \lambda_3 -g_Y \sin\theta_W Y\right)Z
 +eQ\gamma 
 +\left(\frac{g}{2}\sin\theta \lambda_8 +g'\cos\theta Y'\right)Z' , 
\end{align}
where $Q={\rm diag}(0,-1,-1), e=g_Y \cos\theta_W$.
The Weinberg angle $\theta_W$ in this case is defined by 
\begin{equation}
 \sin^2\theta_W
\equiv
 \frac{3}{4+3g^2/g'^2}, 
\end{equation}
which shows that the correct Weinberg angle is obtained 
 by choosing a free parameter $U'(1)$ gauge coupling $g'$ appropriately.

%%%%%%%%%%%%%%%%
\subsection{Lepton flavor mixing}
%%%%%%%%%%%%%%%%
In the gauge-Higgs unification, it is not trivial to generate the flavor mixing 
 since Yukawa coupling is originated from the gauge coupling which is flavor universal. 
In the paper \cite{KK}, it was proposed that the flavor mixing in the context of gauge-Higgs unification 
 can be generated in a situation that both of the bulk and brane mass terms are present, 
 because these masses cannot be simultaneously diagonalized in a flavor space. 
The mechanism was applied to the quark sector in \cite{KK, DD, BB}. 
In this subsection, we apply the mechanism to the lepton sector.  
% namely Maki-Nakagawa-Sakata (MNS) mixing matrix in our model.
As was seen before, 
 we have two lepton doublets $L_{3L}^{(0)}$ and $L_{6L}^{(0)}$ per a generation, 
 we must make one of the linear combination of them massive 
 by introducing the brane-localized mass terms 
 and the brane-localized fermions with charges conjugate to leptons $\bar{l}_R^i$
 as was done in the quark sector. 
Here is such mass terms localized at $y=0$. 
\begin{align}
 \mathcal{L}_{\rm BM}
=&
 \int_{-\pi R}^{\pi R} \d y\sqrt{2\pi R}\delta(y)
 \bar L_R^i(x)[\eta_{ij}L_{3L}^j(x,y)+\lambda_{ij}L_{6L}^j(x,y)]
 \nt\\
\supset&
 \sqrt{2\pi R}\bar L_R^i(x)(\eta_{ij}f_L^j(0),\lambda_{ij}f_L^j(0))
 \left(\begin{array}{c}L_{3L}^{(0)}\\L_{6L}^{(0)}\end{array}\right)
 \nt\\
=&
 \sqrt{2\pi R}\bar L_R'
 [m_\text{diag},0]
 \left(\begin{array}{c}L_{\rm H}\\ L_{\rm SM}\end{array}\right) , 
\end{align}
where $f^i_L(0)$ is a zero mode function of the $i$-th generational lepton doublet evaluated at $y=0$. 
It shows that the $L_{\rm H}$ become massive and decouple from the low-energy effective theory.
The other lepton doublet $L_{\rm SM}$ which corresponds to the standard model leptons is given by
\begin{align}
\label{U3U4}
 \left(\begin{array}{c}L_{3L}^{(0)}\\L_{6L}^{(0)}\end{array}\right)
= 
 U
 \left(\begin{array}{c}L_{\rm H}\\L_{\rm SM}\end{array}\right)
=
 \left(\begin{array}{cc}U_1&U_3\\U_2&U_4\end{array}\right)
 \left(\begin{array}{c}L_{\rm H}\\L_{\rm SM}\end{array}\right)
=
 \left(\begin{array}{c}
 U_1L_{\rm H}+U_3L_{\rm SM}\\U_2L_{\rm H}+U_4L_{\rm SM}
 \end{array}\right)
\to
 \left(\begin{array}{c}
 U_3L_{\rm SM}\\U_4L_{\rm SM}
 \end{array}\right),
\end{align}
where $U$ stands for $6 \times 6$ unitarity matrix.
Then the lepton Yukawa couplings %of the above $L_\text{SM}$ 
 is obtained from their gauge interactions as %combining these expressions as follows:
\begin{align}
 \mathcal{L}_{\rm Lepton~Yukawa}
\supset&
 -\frac{g}{2}\bar e H L_3
 +\frac{g}{\sqrt{2}}\bar \nu H^{\rm T} (i\sigma_2)^\ast L_6
 +({\rm h.c.})
 \nt\\
\supset&
  -\frac{g}{2}\frac{h}{\sqrt{2\pi R}}\bar e_{R}^{i(0)} U_3^{ij}I_{RL}^{i(00)}e_L^{j(0)}
 +\frac{g}{\sqrt{2}}\frac{h}{\sqrt{2\pi R}}\nu_{R}^{i(0)}
  U_4^{ij}I_{RL}^{i(00)}\nu_L^{j(0)}
 +({\rm h.c.}),
\end{align}
where $H$ is a SM Higgs doublet and $h$ is its neutral component. 
$I_{RL}^{i(00)}$ is an overlap integral of lepton zero mode functions;
\begin{equation}
I_{RL}^{i(00)}= \int_{-\pi R}^{\pi R} \d y f_L^i(y) f_R^i(y) .
\end{equation}
The mass eigenstates  
\begin{equation}
\begin{cases}
&\tilde e_R^i=V_{eR}^{ij}e_{R}^{j(0)},~~ \tilde e_L^i=V_{eL}^{ij}e_{L}^{j(0)},\\
&\tilde \nu_R^i=V_{\nu R}^{ij}\nu_{R}^{j(0)},~~
\tilde \nu_L^i=V_{\nu L}^{ij}\nu_{L}^{j(0)},
\end{cases}
\end{equation}
are obtained by the ordinary bi-unitary transformations as
\begin{equation}
\begin{cases}
\frac{1}{2}V_{eR}I_{RL}^{(00)}U_3 V_{eL}^\dag=\frac{M_e}{M_W},\\
\frac{1}{\sqrt{2}}V_{\nu R}I_{RL}^{(00)}U_4 V_{\nu L}^\dag=\frac{M_\nu}{M_W},
\end{cases}
\end{equation}
where the eigenvalues are $M_e=\text{diag}(m_e,m_\mu,m_\tau), M_\nu=\text{diag}(m_{\nu_1},m_{\nu_2},m_{\nu_3})$.
The model parameters are fitted by the above lepton masses and 
 Maki-Nakagawa-Sakata (MNS) mixing matrix $V_{\rm MNS}$ obtained by
$V_{\rm MNS} = V_{eL}^\dag V_{\nu L}.$
The results of numerical calculation are listed in the appendix \ref{modelparameter}.

%%%%%%%%%%%%%%%%%
\subsection{Lepton flavor violation}
%%%%%%%%%%%%%%%%%
In the previous subsection, we discussed how the lepton flavor mixing occurs.
These flavor mixings cause the FCNC vertices with the KK mode gauge bosons 
 since the gauge couplings are flavor dependent.
From now, we focus on the neutral current sector and derive the FCNC interactions 
 which are necessary to calculate $\mu \to 3e$ and $\mu \to e$ conversion processes.
Extracting the photon and $Z$-boson interactions, we obtain the following expression.
\begin{align}
 \mathcal L_{\rm l}
\supset&
 \bar l_3^i\left[\frac{g}{2}\lambda^a A_M^a +g'Y'A'_M\right]\Gamma^Ml_3^i
 +{\rm Tr} \bar l_{\bar 6}\left[-g \lambda^a A_M^a +g'Y'A'_M\right]\Gamma^M l_{\bar 6}
 \nt\\
\supset&
 e[-\bar e_3 \gamma^\mu e_3-\bar e_6 \gamma^\mu e_6 -\bar e \gamma^\mu e 
  ]\gamma_\mu
 \nt\\
&+
 \frac{1}{2}g\cos\theta_W \left[%(1+\tan^2\theta_W)\bar \nu_3 \gamma^\mu \nu_3
 %+
 (-1+\tan^2\theta_W) \bar e_3 \gamma^\mu e_3 +2\tan^2\theta_W \bar e \gamma^\mu e\right]Z_\mu
 \nt\\
&
 +g\cos\theta_W
 [
  \frac{1}{2}(-1+\tan^2\theta_W)\bar e_6\gamma^\mu e_6 
  ]Z_\mu, 
\end{align}
where $e_3$ and $e_6$ stand for the electron which are included in the $L_3 $ and $ L_6$.
Integrate out the fifth extra dimension, 
 four dimensional gauge interactions are obtained as follows; 
\begin{align}
 \mathcal{L}_\text{GI}^\text{4D}
=&
 \int_{-\pi R}^{\pi R} \d y ~ \mathcal {L}_\text{l}
 \nt \\
\supset &
 -e \sum_{n} \left\{ 
 \bar{\tilde e} \gamma^\mu 
 \left[
 LV_{eL}^\dag (U_3^\dag I_n^{L} U_3+U_4^\dag I_n^{L} U_4)V_{eL}
 +RV_{eR}^\dag I_n^{L} (-1)^n V_{eR}
 \right]\tilde e \gamma_\mu^{(n)}
 \right.  \nt\\
 & \left. 
 +\frac{1}{2}g\cos\theta_W \bar {\tilde e}\gamma^\mu
 \left[
 L(-1+\tan^2 \theta_W) V_{eL}^\dag (U_3^\dag I_n^{L} U_3 +U_4^\dag I_n^{L} U_4)V_{eL} 
 \right. \right. \nt \\
& \left. \left. 
+R2\tan^2 \theta_W %V_{eR}^\dag 
V_{eR}^\dag I_n^{L} (-1)^n V_{eR}
 \right]\tilde e Z_\mu^{(n)} 
 \right\},
\end{align}
where the chiral projection operators are used $L=(1+\gamma_5)/2$ and $R = (1-\gamma_5)/2$. 
The $I_n^{L}$ are the integration of the profiles of KK mode gauge boson and zero mode leptons
and their explicit form will be given in the next section. 
The obtained vertices indicate that the FCNC process appears by the neutral KK $Z$ and photon exchange at tree level 
 as in the case of KK gluon exchange in the quark sector \cite{KK, DD, BB}.

%%%%%%%%%%%%%%%%%%%%%
\section{Lepton flavor violation processes}
%%%%%%%%%%%%%%%%%%%%%
We are interested in several lepton flavor violation processes 
 such as $\mu\to e \gamma, \mu\to 3e$ and $\mu \to e$ conversion in atomic nuclei 
 which are focused in the ILC experiments.   
In this paper, we concentrate on the $\mu \to 3e$ and $\mu \to e$ conversion for the following reason.
The $\mu \to e \gamma$ process appears at one loop contributions 
 in contrast with the others arise from tree-level contributions,
therefore we expect that the $\mu \to e \gamma$ process is suppressed compared with the other two processes.

%%%%%%%%%%%%%%%
\subsection{$\mu \to 3e$ process}
%%%%%%%%%%%%%%%
Now we are ready to calculate a lepton flavor violation (LFV) process: $\mu \to 3e$ 
 which is one of the main issue of the ILC experiment. 
The recent experiment \cite{3e} puts a upper bound for $\mu \to 3e$
\begin{equation}
{\rm Br}(\mu^+ \to e^+e^-e^+)=\frac{\Gamma(\mu^+\to e^+e^-e^+)}{\Gamma_{\rm total}}< 1.0\times 10^{-12}.
\end{equation}
Since the ordinary muon decay process $\mu\to \nu_\mu e \bar \nu_e$ is dominated within the total decay width of muon $\Gamma_{\rm total}$,
it can be replaced with $\Gamma(\mu\to \nu_\mu e \bar \nu_e)$ in a good approximation.

To calculate the decay width $\Gamma(\mu \to e^-e^+ e^-)$, 
we first consider the general formulae of $\mu\to 3e$ process.
The generalized vertex functions in our model are 
\begin{align}
 \begin{array}{c}
 \includegraphics{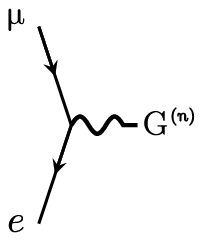}
 \end{array}
&
=\gamma^\mu(A_n^{GL} L +A_n^{GR}R), 
\\
 \begin{array}{c}
 \includegraphics{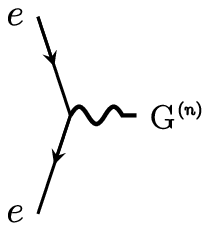}
 \end{array}
&
=\gamma^\mu(B_n^{GL} L +B_n^{GR}R), 
\\
 \begin{array}{c}
 \includegraphics{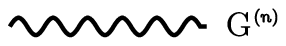}
 \end{array}
&
=\frac{\eta_{\mu\nu}}{q^2-(M_n^G)^2}.
\end{align}
%where chiral projection operators are defined by $L=(1-\gamma_5)/2$ and $R=(1+\gamma_5)/2$.
We adopt the abbreviation $G$ in the superscript of $A, B$ as an intermediate neutral gauge boson: $G=\gamma, Z$.\footnote{
Note that we must take into account the anomalous $U(1)$ gauge boson $Z'$ in $G$, 
 however, we can neglect the contributions for the following reason.
The anomaly in this model appears at the fixed points, 
 so that it yields the brane localized mass term of the anomalous gauge boson \cite{PSW}. 
Since the profile of the anomalous gauge boson behaves as $\sin |y|$, 
 the couplings to the zero mode leptons are expected to be small due to the locality of the zero mode leptons.}
The photon and Z-boson vertex functions are found as follows.  
\begin{eqnarray}
%\begin{cases}
A_n^{\gamma L}
  &=& -e \left[V_{eL}^\dag (U_3^\dag J_n^{L}U_3 +U_4^\dag J_n^{L}U_4)V_{eL}\right]_{\mu e}
%  \\
%  &\sim& -e(5.04 \times 10^{-6} J_L^\tau-6.08 \times 10^{-7} J_L^\mu -4.43\times10^{-6} J_L^e)
, 
  \\
A_n^{\gamma R}
  &=& -e[V_{eR}^\dag J_n^{L}(-1)^n V_{eR}]_{\mu e}
%  \\
%  &\sim& -e(2.70 \times 10^{-15}J_R^\tau+2.14 \times 10^{-8}J_R^\mu -2.14 \times10^{-8} J_R^e)
, 
  \\
B_n^{\gamma L}
  &=& -e \left[V_{eL}^\dag (U_3^\dag J_n^{L}U_3 +U_4^\dag J_n^{L}U_4)V_{eL}\right]_{e e}
%  \\
%  &\sim& -e(1.18 \times 10^{-6}J_L^\tau+4.84\times 10^{-7}J_L^\mu +1.00 J_L^e)
, 
  \\
B_n^{\gamma R}
  &=& -e[V_{eR}^\dag J_n^{L}(-1)^n V_{eR}]_{e e}
%  \\
%  &\sim& -e(3.33 \times 10^{-18} J_R^\tau +4.60 \times 10^{-16} J_R^\mu+1.00 J_R^e)
, 
  \\
A_n^{Z L}
  &=& \frac{1}{2}g\cos\theta_W(-1+\tan^2 \theta_W) \left[V_{eL}^\dag (U_3^\dag I_n^{L}U_3 +U_4^\dag I_n^{L}U_4)V_{eL}\right]_{\mu e} 
%  \sim 0.638 A_n^{\gamma L}
, 
  \\
A_n^{Z R}
  &=& g\cos\theta_W\tan^2\theta_W[V_{eR}^\dag I_n^{L}(-1)^n V_{eR}]_{\mu e}
%  \sim -0.548 A_n^{\gamma R}
, 
  \\
B_n^{Z L}
  &=& \frac{1}{2}g\cos\theta_W(-1+\tan^2 \theta_W) \left[V_{eL}^\dag (U_3^\dag I_n^{L}U_3 +U_4^\dag I_n^{L}U_4)V_{eL}\right]_{e e}
% \sim 0.638 B_n^{\gamma L}
, 
  \\
B_n^{Z R}
  &=& g\cos\theta_W\tan^2\theta_W[V_{eR}^\dag I_n^{L}(-1)^n V_{eR}]_{e e}
  %\sim -0.548 B_n^{\gamma R}
  .  
%\end{cases}
\end{eqnarray}
The overlap integrals of mode functions between the zero mode leptons and the KK photons (KK Z-bosons) are 
\begin{equation}
I_n^{L}=\int_{-\pi R}^{\pi R} \d y \, (f_L^i)^2C_n=J_n^i,~ 
I_n^{R}=\int_{-\pi R}^{\pi R} \d y \, (f_R^i)^2C_n=(-1)^n J_n^i ,(i=e,\mu,\tau) 
\end{equation}
where $C_n$ and $f_L^i$ are the mode functions of gauge bosons and zero mode fermions 
which are defined by the section \ref{Themodel}
%$C_n=\cos\frac{ny}{R}/\sqrt{\pi R},f_L^i=\sqrt{M^i/(1-\e^{-2\pi RM^i})}\e^{-M^i |y|}$ 
and 
\begin{equation}
J_n^i
  =\frac{1}{\sqrt{\pi R}}\frac{(2RM^i)^2}{1-\e^{-2\pi RM^i}}
   \frac{1-(-1)^n\e^{-2\pi RM^i}}{n^2+(2M^iR)^2}.
\end{equation}
The numerical values used in the above calculations are found in appendix A. 

By using these expressions, 
 we can write down the amplitude $\mathcal M$ of $\mu \to 3e$ process as
\begin{equation}
\mathcal M=
\sum_G\sum_{n=1}^\infty
\bar e(p_1)\gamma^\mu (A_n^{GL}L+A_n^{GR}R)\mu (k)
\bar e(p_3)\gamma_\mu (B_n^{GL}L+B_n^{GR}R)e(p_2)
\frac{1}{(k-p_1)^2-(M_n^G)^2}.
\end{equation}
In this calculation, we treat the spin of the initial state ($\mu$) and the final state ($e$) 
 as \lq\lq spin average \rq\rq and \lq\lq spin sum\rq\rq, respectively. 
After the straightforward lengthy algebra, 
 we obtain the partial decay width of $\Gamma(\mu \to 3e)$ as
\begin{align}
 \Gamma(\mu \to 3e)
=& 
 \frac{m_\mu}{128\pi^3}
 \left[ -2m_e^3 m_\mu S_1 - \frac{1}{48} m_e m_\mu^3 S_2 
 + \frac{1}{3} m_e^2 m_\mu^2 S_3 + \frac{2}{15} m_\mu^4 (S_4+S_5) 
 \right]
\end{align} 
where the mode summations are defined as 
\begin{align}
&\dis S_1=\sum_{G,G'}\sum_{n,n'}\frac{1}{(M_n^G)^2(M_{n'}^{G'})^2}
 (A_n^{GR}A_{n'}^{G'L\ast}+A_n^{GL}A_{n'}^{G'R\ast})
 (B_n^{GR}B_{n'}^{G'L\ast}+B_n^{GL}B_{n'}^{G'R\ast}),
 \\
&\dis S_2=\sum_{G,G'}\sum_{n,n'}\frac{1}{(M_n^G)^2(M_{n'}^{G'})^2}
 (A_n^{GR}A_{n'}^{G'L\ast}+A_n^{GL}A_{n'}^{G'R\ast})
 (B_{n'}^{G'L\ast}B_n^{GL}+B_n^{GR}B_{n'}^{G'R\ast}),
\\
&\dis S_3=\sum_{G,G'}\sum_{n,n'}\frac{1}{(M_n^G)^2(M_{n'}^{G'})^2}
 (A_n^{GL}A_{n'}^{G'L\ast}+A_n^{GR}A_{n'}^{G'R\ast})
 (B_{n'}^{G'R\ast}B_n^{GL}+B_n^{GR}B_{n'}^{G'L\ast}),
\\
&\dis S_4=\sum_{G,G'}\sum_{n,n'}\frac{1}{(M_n^G)^2(M_{n'}^{G'})^2}
 (A_n^{GL}A_{n'}^{G'L\ast}B_{n}^{GR}B_{n'}^{G'R\ast}+A_n^{GR}A_{n'}^{G'R\ast}B_{n}^{GL}B_{n'}^{G'L\ast}),
\\
&\dis S_5=\sum_{G,G'}\sum_{n,n'}\frac{1}{(M_n^G)^2(M_{n'}^{G'})^2}
 (A_n^{GL}A_{n'}^{G'L\ast}B_{n}^{GL}B_{n'}^{G'L\ast}+A_n^{GR}A_{n'}^{G'R\ast}B_{n}^{GR}B_{n'}^{G'R\ast}).
\end{align}
These mode summations are very complicated and we calculated them numerically as
\begin{align}
&S_1=2.38 \times 10^{-16} g^4 R^4
   \sim  4.45 \times{10}^{-15}\,{e}^{4} R^4,\\
&S_2=-3.04 \times 10^{-17} g^4 R^4
   \sim -5.69 \times{10}^{-16}\,{e}^{4} R^4,\\
&S_3=-9.68 \times 10^{-15} g^4 R^4
   \sim -1.81 \times{10}^{-13}\,{e}^{4} R^4,\\
&S_4=3.38 \times 10^{-16} g^4 R^4
   \sim  6.33 \times{10}^{-15}\,{e}^{4} R^4,\\
&S_5=6.92 \times 10^{-14} g^4 R^4
   \sim  1.29 \times{10}^{-12}\,{e}^{4} R^4.
\end{align}
We note that the $SU(2)$ gauge coupling $g$ is replaced by the $U(1)_{em}$ gauge coupling $e$ 
 through the relation $e=g\sin\theta_W$. 
%where $g$ is the SU(2) gauge coupling.
%(The mode  summations are cut at $n=20$)
We finally obtain the partial decay width of $\mu \to 3e$ in the gauge-Higgs unification
\begin{equation}
\Gamma(\mu \to 3e) \sim 4.45 \times 10^{-13} (R M_W)^4,
%\Rightarrow R\geq 0.81 M_W
\end{equation}
which tells us the lower bound for the compactification scale $R^{-1} \geq 0.81 M_W \sim 65\rm GeV$. 

%%%%%%%%%%%%%%%
\subsection{$\mu \to e$ conversion}
%%%%%%%%%%%%%%%

In this subsection, we turn to another process of lepton flavor violation at tree level in gauge-Higgs unification, 
 e.g. $\mu \to e$ conversion in nuclei. 
The four dimensional effective Lagrangian describing $\mu \to e$ conversion process is given 
 by the following 4-Fermi interactions \cite{CGTT,CN}
\begin{equation}
{\cal L}_{\mu \to e}\supset  \frac{G_F}{\sqrt{2}} \left[ \bar{e} \gamma^\mu (v - a \gamma_5 ) \mu \sum_{q=u,d} \bar{q} (v^q - a^q \gamma_5) q \right],
\label{mueconv}
\end{equation}
where the lepton current part is the same form used in the calculations of $\mu \to 3e$ with factors 
\begin{equation}
v = \frac{1}{2} (A_n^{GL} + A_n^{GR}), \quad a = \frac{1}{2} (A_n^{GL} - A_n^{GR})~(G = \gamma_\mu, Z_\mu). 
\end{equation}
The four dimensional effective neutral current of the quark sector in (\ref{mueconv}) can be found 
 from the photon and the Z-boson currents of the model in \cite{BB}.  
\bea
{\cal L}_{{\rm NC}}^{4D} &=& 
\bar{u} \gamma^\mu (LB_n^{\gamma Lu} + RB_n^{\gamma Ru}) u \gamma^{(n)}_\mu
+ \bar{d} \gamma^\mu (LB_n^{\gamma Ld} + RB_n^{\gamma Rd}) d \gamma^{(n)}_\mu 
\nonumber \\
&& + \bar{u} \gamma^\mu (LB_n^{Z Lu} + RB_n^{Z Ru}) u Z^{(n)}_\mu
+ \bar{d} \gamma^\mu (LB_n^{Z Ld} + RB_n^{Z Rd}) d Z^{(n)}_\mu,
\label{effectivelagrangian}
\eea   
where
\bea
%\begin{cases}
B_n^{\gamma Lu} &=& \frac{2}{3}e [V_{uL}^\dag (U^\dag_{q3} I_n^L U_{q3} + U^\dag_{q4} I_n^L U_{q4}) V_{uL}]_{uu}, \\
B_n^{\gamma Ru} &=& \frac{2}{3}e [V_{uR}^\dag I_n^L (-1)^n V_{uR}]_{uu}, \\
B_n^{\gamma Ld} &=& -\frac{1}{3}e [V_{dL}^\dag (U^\dag_{q3} I_n^L U_{q3} + U^\dag_{q4} I_n^L U_{q4}) V_{dL}]_{dd}, \\
B_n^{\gamma Rd} &=& -\frac{1}{3}e [V_{dR}^\dag I_n^L (-1)^n V_{dR}]_{dd}, \\
B_n^{Z Lu} &=& \frac{g}{\cos \theta_W} \left( \frac{1}{2} - \frac{2}{3} \sin^2 \theta_W \right) [V_{uL}^\dag (U^\dag_{q3} I_n^L U_{q3} 
+ U^\dag_{q4} I_n^L U_{q4}) V_{uL}]_{uu}, \\
B_n^{Z Ru} &=&\frac{g}{\cos \theta_W} \left( - \frac{2}{3} \sin^2 \theta_W \right) [V_{uR}^\dag I_n^L (-1)^n V_{uR}]_{uu}, \\
B_n^{Z Ld} &=& \frac{g}{\cos \theta_W} \left( \frac{1}{2} + \frac{1}{3} \sin^2 \theta_W \right) [V_{dL}^\dag (U^\dag_{q3} I_n^L U_{q3} + U^\dag_{q4} I_n^L U_{q4}) V_{dL}]_{dd}, \\
B_n^{Z Rd} &=& \frac{g}{\cos \theta_W} \left(  \frac{1}{3} \sin^2 \theta_W \right) [V_{dR}^\dag I_n^L (-1)^n V_{dR}]_{dd}.
%\end{cases}
\eea
The $U_{q3} $ and $U_{q4}$ are the mixing matrices representing 
 how much the $Q_\text{SM}$ is included in the $Q_3,Q_6$ and $Q_{15}$, 
 which correspond to the (\ref{U3U4}) in the lepton sector.  
$V_{uL(R)}$ and $V_{dL(R)}$ are the unitarity matrices diagonalizing up-type and down-type yukawa couplings, respectively. 
The parameters in the quark sectors are fitted similarly to the leptons and they are summarized in the appendix \ref{modelparameter}.  
Then the factors in the eq. (\ref{mueconv}) can be read off from eq. (\ref{effectivelagrangian}) as 
\beq
v^q=\frac{1}{2} (B^{GLq}_n +B^{GRq}_n)\ ,\  a^q=\frac{1}{2} (B^{GLq}_n -B^{GRq}_n).
\eeq
The $\mu \to e $ conversion rate in the light nuclei is precisely discussed in the paper \cite{CGTT,CN}
\begin{equation}
\label{conversion}
 B_\text{conv} 
= 
 \frac{2p_e E_e F_p^2 m_\mu^3 \alpha^3 Z_\text{eff}^2}{\pi^2 Z\Gamma_\text{capt}}
 \big[
  |(v^\gamma-a^\gamma)Q_N^\gamma + (v^Z-a^Z)Q_N^Z |^2 +|(v^\gamma+a^\gamma)Q_N^\gamma + (v^Z+a^Z)Q_N^Z |^2 
  \big],
\end{equation}
where $Q_N^\gamma = [v^u (2Z+N)+v^d (2N+Z)]|_{G=\gamma}$ and $Q_N^Z = [v^u (2Z+N)+v^d (2N+Z)]|_{G=Z}$.
$Z$ and $N$ represent the atomic number and neutron number of the target nuclei.
The most sensitive experimental result is the case for the ${}^{48}_{22}\rm Ti$. 
The parameters in the above expressions are
\begin{equation}
E_e\sim p_e \sim m_\mu\ ,~ F_p\sim 0.55\ ,~ Z_\text{eff}\sim 17.61\ ,~ \Gamma_\text{capt}\sim 2.6\times 10^6 [{\rm s^{-1}}]\ .
\end{equation}
Putting them into eq. (\ref{conversion}) and summing up the KK modes of internal gauge bosons, 
 we obtain the following results
\begin{equation}
B_\text{conv}  =
\begin{cases}
2.89\times 10^{-4} R^4  ~~&(\text{for the case }R_u={\bf 1}_{3\times 3}) \ ,
\\
1.46\times 10^{-4} R^4  ~~&(\text{for the case }R_d={\bf 1}_{3\times 3}) \ ,
\end{cases}
\end{equation}
which lead to the lower bounds of compactification scale as
\begin{equation}
R^{-1}\geq
\begin{cases}
147.5{\rm GeV}  ~~&(\text{for the case }R_u={\bf 1}_{3\times 3}) \ ,
\\
69.89{\rm GeV} ~~&(\text{for the case }R_d={\bf 1}_{3\times 3}) \ ,
\end{cases}
\end{equation}
from the experimental data $\rm{Br}(\mu\to e)_{\rm Ti} < 6.1\times 10^{-13}$ \cite{Wintz:1996va}.

%%%%%%%%%%
\section{Summary}
%%%%%%%%%%
In this paper, we have discussed the lepton flavor violation within the gauge-Higgs unification model.
Yukawa couplings are essentially universal since they are gauge coupling in this scenario.
The flavor violation is achieved by the interplay between the fermion bulk mass terms 
 localizing the leptons and quarks at fixed points 
 and the brane localized mass terms removing the extra massless fermions. 
We have no flavor violation on the neutral gauge interaction in the zero mode sector 
 due to the universality of the gauge coupling, 
 but the tree level FCNC vertex appears in the KK mode gauge boson sector 
 since the gauge interactions in the KK mode sector are found to be flavor dependent.  

These tree level FCNC interactions may give rise to the large lepton flavor violation processes 
 such as the $\mu \to 3e$ decay processes and the $\mu \to e$ conversion in atomic nuclei, 
 which is one of the main purpose of the future ILC experiment. 
%We calculate the branching ratio of $\mu \to 3e$ process.
Though these process takes place at tree level, they are rather small against our expectation 
 and we obtain the lower bound of compactification scale as $1/R\geq \mathcal O(M_W)$.

The reason why the lepton violations are suppressed is that the lepton flavor symmetry is almost conserved in this model. 
It is the general feature of this model that the differences between the eigenstates of bulk masses and brane masses of fermions 
 is the only source of the flavor violation in contrast to the other models 
 such as the supersymmetric model, 
 in which the large flavor violation other than Yukawa coupling is in general present in soft SUSY breaking parameters.
If the neutrino masses are degenerate, 
%for examples the bulk masses of the leptons are degenerate, 
 the lepton flavor violation completely disappears.
Taking into account this fact, 
 the final results roughly receive the suppression factor as $\Delta m_\nu^2 R^2$ reflecting the flavor structure, 
  where $\Delta m_\nu^2$ denotes the differences of neutrino mass squared. 
For example, the $\mu \to 3e$ processes which was argued in the main text are naively estimated as
\begin{equation}
\text{Br} (\mu\to 3e ) \sim \left(m_W R \right)^4\leq 10^{-12}. 
\end{equation}
Without such kind of suppressions, we find more stringent bound $1/R\sim 10^3 M_W$ than our result. 
However, if the factor $\Delta m_\nu^2 R^2$ is taken into account, 
 we obtain the branching ratio as  
\begin{equation}
\text{Br} (\mu \to 3e ) \sim \Delta m_\nu^2 R^2 \left(m_W R \right)^4\leq 10^{-12} 
\end{equation}
which gives the lower bound $1/R \sim M_W$ with the observed neutrino mass $\Delta m_\nu \leq \mathcal O (10 \rm MeV)$. 
This is the physical reason that the lepton flavor violation considered in this paper is unexpectedly suppressed 
 although it happens even at tree level. 
From this observation, it is very important to study a loop-induced process $\mu \to e \gamma$ in the gauge-Higgs unification 
 since it is expected to provide a stronger bound on model parameters. 
This issue will be left for our future work.

%%% Acknowledgments %%%%%%%%%%%%%%%%%%%%%%%%%%%
\subsection*{Acknowledgments} 
This work was supported in part by the Grant-in-Aid for Scientific Research 
of the Ministry of Education, Science and Culture, No.~24540283 (N.M.). 
%%%%%%%%%%%%%%%%%%%%%%%%%%%%%%%%%%%%%%

\appendix

\section{Model parameters}
\label{modelparameter}
In this appendix A, the details of parameters used in the calculations are summarized. 

\subsection{lepton sector}

Yukawa coupling and MNS matrix in the lepton sector are given by 
\begin{align}
%\begin{cases}
&\hat Y_e =  V_{eR}^\dag I_{RL}^{(00)}U_3 V_{eL}, \quad 
\hat Y_\nu =V_{\nu R}^\dag \sqrt{2} I_{RL}^{(00)} U_4 V_{\nu L}, \\
%\end{cases}, \\
&V_{\rm MNS} = V_{eL}^\dag V_{\nu L}
= 
\left(
\begin{array}{ccc}
c_{12} c_{13} & s_{12}c_{13} & s_{13} \\
-s_{12}c_{23} -c_{12}s_{23}s_{13} & c_{12}c_{23} -s_{12}s_{23}s_{13} & s_{23}c_{13} \\
s_{12}s_{23}-c_{12}c_{23}s_{13} & -c_{12}s_{23}-s_{12}c_{23}s_{13} & c_{23}c_{13} \\
\end{array}
\right),
\end{align}
where 
\begin{equation}
U_4=R_\nu \left[\begin{array}{ccc}a_{l1}&0&0\\ 0&a_{l2}&0\\ 0&0&a_{l3}\end{array}\right],
U_3=R_e \left[\begin{array}{ccc}\sqrt{1-a_{l1}^2}&0&0\\ 0&\sqrt{1-a_{l2}^2}&0\\ 0&0&\sqrt{1-a_{l3}^2}\end{array}\right]
,
\end{equation}
\begin{align}
&
R_\nu = \left[\begin{array}{ccc}
	1 &0 &0 \\
	0&\cos\theta_{l2}'&\sin\theta_{l2}' \\
	0&-\sin\theta_{l2}'&\cos\theta_{l2}'
	\end{array}\right]
         \left[\begin{array}{ccc}
	\cos\theta_{l3}' &0 &\sin\theta_{l3}' \\
	0&1&0 \\
	-\sin\theta_{l3}'&0&\cos\theta_{l3}'
	\end{array}\right]
        \left[\begin{array}{ccc}
	\cos\theta_{l1}' &-\sin\theta_{l1}'&0 \\
	\sin\theta_{l1}'&\cos\theta_{l1}'&0 \\
	0&0&1
	\end{array}\right]
	,\\
&
R_e = \left[\begin{array}{ccc}
	1 &0 &0 \\
	0&\cos\theta_{l2}&\sin\theta_{l2} \\
	0&-\sin\theta_{l2}&\cos\theta_{l2}
	\end{array}\right]
         \left[\begin{array}{ccc}
	\cos\theta_{l3} &0 &\sin\theta_{l3} \\
	0&1&0 \\
	-\sin\theta_{l3}&0&\cos\theta_{l3}
	\end{array}\right]
        \left[\begin{array}{ccc}
	\cos\theta_{l1} &-\sin\theta_{l1}&0 \\
	\sin\theta_{l1}&\cos\theta_{l1}&0 \\
	0&0&1
	\end{array}\right]
	,
	\\
&
I_{RL}^{(00)}=\text{diag} [b_1^l,b_2^l,b_3^l], \left(b_i^l=\frac{\pi RM^i_l}{\sinh \pi RM^i_l}\right), \\
&s_{ij} \equiv \sin \phi_{ij}, \quad c_{ij} \equiv \cos \phi_{ij}.  
\end{align}
$V_{eL(R)}$ are the unitary matrices diagonalizing the matrices $\hat{Y}_e^\dag \hat{Y}_e (\hat{Y}_e \hat{Y}_e^\dag)$ 
and $V_{\nu L(R)}$ are the unitary matrices diagonalizing the matrices $\hat{Y}_\nu^\dag \hat{Y}_\nu (\hat{Y}_\nu \hat{Y}_\nu^\dag)$.  
In the MNS matrix, CP phase is neglected since CP violation is not an issue in this paper. 

The input parameters for physical observables we should fit are \cite{PDG}
\begin{align}
&\hat m_{\nu_e}=(2.00 \times 10^{-9})/m_W, \quad 
\hat m_{\nu_\mu}=(1.90 \times 10^{-4})/m_W, \quad 
\hat m_{\nu_\tau}=0.0182 /m_W,\\
&\hat m_e=(5.11 \times 10^{-4})/m_W, \quad 
\hat m_\mu=0.106 /m_W, \quad 
\hat m_\tau=1.78 /m_W, \\
& m_{\nu_e} \sim 2 {\rm eV}, \quad m_{\nu_\mu} \sim 190 {\rm keV}, \quad m_{\nu_\tau} \sim 18.2 {\rm MeV}, \\
& \sin^2 \phi_{12} = 0.306, \sin^2 \phi_{23} =0.42, \sin^2 \phi_{13} =0.021. 
\end{align}
%with $m_W=80.5$GeV. 

%\paragraph{fitting parameter}~
We found a set of numerical solutions reproducing the above physical observables in the special case $R_\nu=I_{3 \times 3}$.
\begin{align}
%&\theta'_{l1}=\theta'_{l2}=\theta'_{l3}=0, \\
%~~
%(R_\nu=I_{3\times 3})
%\end{equation}
%\begin{align}
&
\sin\theta_{l1}= 5.53 \times 10^{-1}, %\quad 
%\\
\sin\theta_{l2}=6.48 \times 10^{-1}, %\quad 
%\\
\sin\theta_{l3}=1.50 \times 10^{-1}, \quad 
\\
&
\theta'_{l1}=\theta'_{l2}=\theta'_{l3}=0, \\
&
a_{l1} = 2.77 \times 10^{-6}, \quad %\\ 
a_{l2} = 1.27 \times 10^{-3}, \quad %\\
a_{l3} = 7.24 \times 10^{-3}, \\
&
b_1^l=6.35 \times 10^{-6}, \quad 
%\\
b_2^l=1.31 \times 10^{-3}, \quad 
%\\
b_3^l=2.21 \times 10^{-2}. 
\end{align}

%\paragraph{Yukawa coupling, mixing matrix}
Then, Yukawa coupling and their mixing matrix in terms of these numerical solutions are listed below. 
%\subparagraph{neutrino}
%~
\begin{align}
 &Y_\nu
=
 \sqrt{2}I_{RL}^{(00)}U_4\nt 
=
 \begin{pmatrix}
 2.48 \times 10^{-11} & 0 & 0\cr
 0 & 2.36 \times 10^{-6} & 0\cr 
 0 & 0 & 2.26 \times 10^{-4}
\end{pmatrix}
%\end{align}
, \\
&
V_{\nu L}=V_{\nu R}={\bf 1}_{3\times 3}, \\
 &Y_e
=
 I_{RL}^{(00)}U_3\nt
= 
\begin{pmatrix}
  5.23 \times 10^{-6} &
 -3.47 \times 10^{-6} &
  9.52 \times 10^{-7}
 \cr
  4.47 \times 10^{-4} &
  9.03 \times 10^{-4} &
  8.41 \times 10^{-4}
 \cr
 -1.00 \times 10^{-2} & 
 -1.05 \times 10^{-2}& 
  1.66 \times 10^{-2}
 \end{pmatrix},
 \\
&V_{eL}
  =\left[
    \begin{array}{ccc}
    0.824 & 0.340 & -0.454 \\
    -0.547 & 0.688 & -0.477 \\
    0.150 & 0.641 & 0.753
    \end{array}
    \right],
    \\
&V_{eR}
  =\left[
    \begin{array}{ccc}
    1.00 & -2.14 \times 10^{-8} & -1.83 \times 10^{-9} \\
    2.14 \times 10^{-8} & 1.00 & -1.48 \times10^{-6} \\
    1.83 \times 10^{-9} & 1.48 \times 10^{-6} & 1.00
    \end{array}
    \right],
\end{align}
\begin{align}
U_3
=
\begin{pmatrix}
0.824 & -0.547 & 0.150 \\
0.340 & 0.688 & 0.641 \\
-0.454 & -0.477 & 0.753
\end{pmatrix}
,\quad 
%\\
U_4
=
\begin{pmatrix}
2.77\,{10}^{-6} & 0 & 0\\
0 & 1.27 \times 10^{-3} & 0\\
0 & 0 & 7.24 \times 10^{-3}
\end{pmatrix}.
\end{align}

%%%%%%%%%%%%%
\subsection{quark sector}
%%%%%%%%%%%%%
As for the quark sector, numerical solutions were studied in detail in \cite{BB}. 
Therefore, only the results are listed.  
Yukawa couplings and their mixing matrices are  
\begin{align}
	\label{cond_U3,U4}
	\left\{
	\begin{aligned}
	  \hat Y_d &= {\rm diag}(\hat m_d,\cdots) = V_{dR}^\dag I_{RL}^{(00)}U_{q3} V_{dL}\\
	  \hat Y_u &= {\rm diag}(\hat m_u,\cdots) = V_{uR}^\dag WI_{RL}^{(00)}U_{q4}V_{uL}
	\end{aligned}
	\right.\quad , \qquad
	  V_{\rm CKM}\equiv V_{dL}^\dag V_{uL} ,
\end{align}
where $W=\text{diag} (\sqrt{2},\sqrt{2},2)$ which is originated form the normalization factor of $\overline {\bf 15}$.
\begin{align}
&I_{RL}^{(00)}=\text{diag} [b_1^q,b_2^q,b_3^q] \left(b_i^q=\frac{\pi RM^i_q}{\sinh \pi RM^i_l}\right), \\
%\begin{align} 
%%% \label{parametrization} %%%%%%%%%%%%%%%%%%%%%%
	\label{parametrization}
	&
	U_{q4}
  = R_u\!
	\left[
	\begin{array}{ccc}
	 a_{q1} & 0 & 0\\[3pt]
	 0 & a_{q2} & 0\\[3pt]
	 0 & 0 & a_{q3}
	\end{array}
	\right], \qquad 
	U_{q3}
  = R_d\!
	\left[
	\begin{array}{ccc}
	 \sqrt{1-a_{q1}^2} & 0 & 0\\[2pt]
	 0 & \sqrt{1-a_{q2}^2} & 0\\[2pt]
	 0 & 0 & \sqrt{1-a_{q3}^2}
	\end{array}
	\right] ,
\end{align}
where $R_u$ and $R_d$ are arbitrary $3\times3$ rotation matrices parametrized as
\begin{subequations}
\begin{align}
	R_u %\label{3gparameterizationRu}
&= 
 \left[\begin{array}{ccc}
	1 &0 &0 \\
	0&\cos\theta_{q2}'&\sin\theta_{q2}' \\
	0&-\sin\theta_{q2}'&\cos\theta_{q2}'
	\end{array}\right]
         \left[\begin{array}{ccc}
	\cos\theta_{q3}' &0 &\sin\theta_{q3}' \\
	0&1&0 \\
	-\sin\theta_{q3}'&0&\cos\theta_{q3}'
	\end{array}\right]
        \left[\begin{array}{ccc}
	\cos\theta_{q1}' &-\sin\theta_{q1}'&0 \\
	\sin\theta_{q1}'&\cos\theta_{q1}'&0 \\
	0&0&1
	\end{array}\right], \\
	R_d
 &=
 \left[\begin{array}{ccc}
	1 &0 &0 \\
	0&\cos\theta_{q2}&\sin\theta_{q2} \\
	0&-\sin\theta_{q2}&\cos\theta_{q2}
	\end{array}\right]
         \left[\begin{array}{ccc}
	\cos\theta_{q3} &0 &\sin\theta_{q3} \\
	0&1&0 \\
	-\sin\theta_{q3}&0&\cos\theta_{q3}
	\end{array}\right]
        \left[\begin{array}{ccc}
	\cos\theta_{q1} &-\sin\theta_{q1}&0 \\
	\sin\theta_{q1}&\cos\theta_{q1}&0 \\
	0&0&1
	\end{array}\right].
\end{align}
\end{subequations}
The two set of numerical solutions were found in \cite{BB}. 
One is a set of solutions with the case $R_u = \bs1_{3\times3}$ where the up-type quark mixings vanish.  
\begin{alignat}{5}
	a_{q1}^2 \approx 0.1023 &\qquad&
	(b_1^q)^2 \approx 4.355\times10^{-9} &\qquad&&
   \sin\theta_{q1} \approx -2.587\times10^{-2}\notag\\
	a_{q2}^2 \approx 0.9887 &\quad , \qquad&
	(b_2^q)^2 \approx 1.302\times10^{-4} &\quad , \qquad&&
   \sin\theta_{q2} \approx 2.224\times10^{-2}~~.\\
	a_{q3}^2 \approx 0.9966 &\qquad&
	&\qquad&&
   \sin\theta_{q3} \approx 2.112\times10^{-4}\notag
\end{alignat}
Another is a set of solutions the case $R_d = \bs1_{3\times3}$ where the down-type quark mixings vanish. 
\begin{alignat}{5}
	a_{q1}^2 \approx 0.0650 &\qquad&
	(b_1^q)^2 \approx 3.973\times10^{-9} &\qquad&&
   \sin\theta_{q1}' \approx 0.6704\notag\\
	a_{q2}^2 \approx 0.9931 &\quad , \qquad&
	(b_2^q)^2 \approx 2.235\times10^{-4} &\quad , \qquad&&
   \sin\theta_{q2}' \approx -3.936\times10^{-2}~~.\\
	a_{q3}^2 \approx 0.9966 &\qquad&
	&\qquad&&
   \sin\theta_{q3}' \approx 1.773\times10^{-2}\notag
\end{alignat}

%%%%%%%%%%%%%%%%%%%%%%%%%%%%%%% bibliography %%%%%%%%%%%%%%%%%%%%%%%%%%%%%%%

\end{document}